\documentclass[preprint,times]{elsarticle}
\usepackage[latin9]{inputenc}
\usepackage{amsmath,amssymb,amsfonts}
\usepackage{esint}
\usepackage{hyperref}
\usepackage{graphicx}
\usepackage{subfig}



\begin{document}

\title{De Broglie relations, gravitational time dilation and weak equivalence principle}

\author[amss,ucas]{Peng Xu\corref{cor}}
\ead{xupeng@amss.ac.cn}

\author[chd]{Li-E Qiang}
\ead{qqllee815@chd.edu.cn}

\address[amss]{Academy of Mathematics and Systems Science, Chinese Academy of Sciences, No.55 Zhongguancun East Road, Haidian District, Beijing, China 100190.}

\address[ucas]{University of Chinese Academy of Sciences, No.19(A) Yuquan Road, Shijingshan District, Beijing, China 100049.}

\address[chd]{Chang'an University, Yanta District, Xi'an, China 710064.}

\cortext[cor]{Corresponding author}

\begin{abstract}
Interplays between quantum physics and gravity have long inspired exciting studies, which
reveal subtle connections between quantum laws and the general notion of curved spacetime.
One important example is the uniqueness of free-falling motions in both quantum and gravitational
physics. In this work, we re-investigate the free-falling motions of quantum test
wave packets that are distributed over weakly curved spacetime backgrounds. Except for the de Broglie
relations, no assumption of priori given Hamiltonians or least actions satisfied by the quantum
system is made.  We find that the mean motions of quantum test wave packets can be deduced
naturally from the de Broglie relations with a generalized treatment of gravitational time
dilations in quantum waves. Such mean motions of quantum test masses in gravitational field
are independent of their masses and compositions, and restore exactly the free-falling or geodesic motions of classical
test masses in curved spacetime. This suggests a novel perspective that weak equivalence principle,
which states the universality of free-fall, may be deeply rooted in quantum physics and be a phenomenon that has
emerged from the quantum world.
\end{abstract}

\maketitle

\section{Introduction}

Quantum theory and general relativity both reveal the unusual properties of the physical
world that depart from our everyday experiences and intuitions.
Interplays between them had long inspired theoretical investigations and exciting studies
of new phenomena. For example, a quantum theory of gravity
that sees the consistent merger of these two fundamental theories
\cite{Kiefer2012} has been a long-sought goal
in theoretical physics. Novel predictions, such as black hole
radiations \cite{Hawking1975,Hawking1976,Wald2001}, had
been drawn from the studies of quantum fields in curved spacetime \cite{Wald1994}.
Ideas of gravitational decoherence  \cite{Bassi2017},
first suggested by  Di\'osi \cite{Diosi1984,Diosi1989} and Penrose \cite{Penrose1996,Penrose2014},
had also shed some new light on the quantum measurement problem.
Recently, gravitational time dilation in quantum interferometry have been
studied in a series of works \cite{Zych2011,Zych2012,Pikovski2015,Pikovski2017},
and their relations with the equivalence principle has brought up active
discussions \cite{Bonder2015,Adler2016,Bonder2016,Pikovski2016,Pang2016,Diosi2017}.

Besides those advances from combining quantum
theory and gravitation, there lies subtle connections and consistencies
between these two seemingly different
perspectives of the physical world. An important example of such consistencies
is the uniqueness of free-falling or inertial motions in both quantum and gravitational physics, which may relate the nature of quantum fluctuations to gravitation
\cite{Smolin1986,Smolin1986a,Smolin1986b}. As summarized by Smolin in \cite{Smolin1986}, that having quantum fluctuations, inertia
and gravitation being the absolutely universal phenomena, the preferred
motions singled out by vanishing the dissipation effects from quantum
fluctuations (spectrum properties of vacuum quantum fluctuations) are exactly the
free-falling or inertia motions considered in classical general relativity.
In the hydrodynamic formulation of quantum mechanics first proposed by Madelung \cite{Madelung1926},
the evolutions of the probability density and
current of a quantum system are governed by the Madelung equations which contain
a quantum potential quadratic in the Planck's constant. The Madelung equations
had then provided the basis for different classical interpretations of quantum mechanics including the
stochastic interpretation due to Nelson \cite{Nelson1966,Nelson1967}.
The uniqueness of inertial motions is also manifested in the stochastic interpretation of quantum
mechanics, that the diffusion
constant or the correlation length of the particle subjected to quantum
fluctuations is inversely proportional to a quantity with the dimension
of mass \cite{Nelson1967,Smolin1986b}, and the equivalence
of such ``quantum mass'' and the inertia mass ensures the linearity
of the (Schrödinger type) equation that is governing the evolution of the
probability density and current of the particle subjected to quantum
fluctuations \cite{Smolin1986,Smolin1986a,Smolin1986b}. Such agreements
between these two seemingly unrelated fields may not be taken merely
as a coincidence, but could possibly be ``the central mystery behind
the question of the relationship between quantum and gravitational
phenomena'' as pointed out by Smolin \cite{Smolin1986}. In this work, we
try to re-investigate such connections through the unique properties of free-falling
or inertial motions. The key result turns out to be that, if in curved spacetime the clock rate
differences or gravitational time dilations in different parts of
a quantum system are taken into account, the universality of inertial
or free-falling (mean) motions of quantum test particles or bodies in gravitational
field may naturally be deduced from the fundamental de Broglie relations.
This leads to a novel perspective that the classical weak equivalence
principle (WEP) may be deeply rooted in quantum physics and be
a phenomenon that has emerged from the quantum world. In deriving
this, after introducing the framework in the next section, we consider a quantum
scalar wave packet propagated in a classical
weakly curved spacetime background, and, except for the de Broglie
relations, no assumption of priori given Hamiltonians or least actions
satisfied by the quantum wave is made.

\section{Weak equivalence principle and local geometry}

The principle of equivalence is the foundation of the geometric view
of gravitation. Summarized in the 1964 Les Houches lecture given by Dicke
\cite{Dicke1964}, the Einstein equivalence principle (EEP)
contains three ingredients: WEP, Lorentz invariance and position
independence of non-gravitational physical laws in local free-falling
frames. WEP states that trajectories of free-falling test particles
or bodies in gravitational fields are independent of their internal
structures and compositions. In the history of the developments of
gravitational theories, from the opening pages of Newton's \emph{Principia}
to Einstein's gedanken experiment of a free-falling elevator, WEP
had played a crucial role. Connections between WEP and the other two
ingredients of EEP are believed to exist, and the Schiff's conjecture,
see \cite{Will2014}, states that any complete and self-consistent
theory of gravity that embodies WEP will necessarily embody the full
EEP. Given the full EEP, it can be further argued that gravitation
must be the phenomena of curved spacetime and satisfy the so called
metric theories \cite{Dicke1964}. WEP, or the universality of free-fall,
is therefore at the very heart of modern theories of gravitation.

If the fundamental notion of curved spacetime is accepted,
in metric theories (including general relativity as the best fit candidate confronted
with the many stringent tests \cite{Will2014,Ni2016}), WEP is fulfilled
due to one basic postulate that matter fields couple to gravity in a universal manner.
For test masses, this means that free-falling motions follow
the shortest paths (geodesic lines) in curved spacetime endowed
with a symmetric metric tensor field $g_{\mu\nu}$ (the indices $\mu,\nu=0,1,2,3$).
Such postulate also imposes absolute structures to these geometric
theories of gravitation, such as the non-dynamical and priori given
coupling constants, see \cite{Dicke1964,Damour2012,Will2014} for detailed
discussions. This had firstly been questioned since 1930s by Dirac and Jordan
\cite{DIRAC1937,Jordan1937,Jordan1939},
which leads to a perspective that WEP may be only an approximation
valid in the low energy realm of gravity
\cite{Damour2012,Will2014}. On the other hand, attempts to construct a full
quantum formulation of WEP are also under the research, see \cite{Zych2018,Anastopoulos2018}
and the references therein. Today, precision tests (including quantum tests
\cite{Viola1996,Zoest2010}) of WEP
and EEP, considered as an approach to discover new physics, are of interest,
and the most updated result from MICROSCOPE has set the upper bound
of the Eötvös-ratio to be about $10^{-15}$ \cite{Touboul2017}. In the following,
we try to tackle this issue from a different angle.

To manifest the potential interweaving between spacetime geometry and
quantum laws, one needs a stage for the low-energy
realm which could bring these two together under common considerations.
In the following, we assume that the velocity $v$ of the test mass
(classical or quantum) is much smaller compared with the speed of
light, and the gravitational field generated by the test mass itself
could be ignored. Units system $c=G=\hbar=1$ is adopted for clarity.
The quantum state of a test particle or a macroscopic quantum system
(such as superfluids) is considered as a square integrable complex
wave function $\Psi(t,\vec{x})$ distributed over a certain spacetime
region. For clarity, spins and other internal degrees of freedom are
not considered in this work. In the low-energy regime, we assume that the
gravitational field or the spacetime curvature is weak in the sense that the scale $L$ of the
wave function $\Psi(t,\vec{x})$ (such as the characteristic width
of Gaussian wave packets) is much smaller compared with the curvature
radius $\mathcal{R}$ in that spacetime region, that $L/\mathcal{R}\ll$1.
Therefore, it is natural to choose Fermi normal coordinate systems
\cite{Nesterov1999} $\{t,\ x^{i}\}$ as reference frames, within
which the spacetime metric components could be expanded in powers
of $\mathcal{O}(\frac{|x|}{\mathcal{R}})$ with $|x|/\mathcal{R}\ll1$.
In a general Riemann manifold, the Fermi normal coordinates system generalized the
classical Riemann normal coordinates (that satisfies the conditions
$g_{\mu\nu}(0)=\eta_{\mu\nu}$ and $\Gamma_{\ \mu\nu}^{\lambda}(0)=0$)
with the connection coefficients $\Gamma_{\ \mu\nu}^{\lambda}$ vanished
along a given curve. With such curve being chosen as a time-like geodesic,
the Fermi normal coordinates system can be physically viewed as the local
frame attached to the observer following that geodesic, and its local tetrad is formed by the
clock and parallel transported rigid rods carried by the observer. Up to
$\mathcal{O}(\frac{|x|^{2}}{\mathcal{R}^{2}})$, the
spacetime metric under the Fermi normal frame can be expanded as \cite{Misner1973,Ni1978}
\begin{eqnarray}
ds^{2} & = & -(1+R_{0i0j}x^{i}x^{j})dt^{2}-(\frac{4}{3}R_{0jik}x^{j}x^{k})dtdx^{i}\nonumber \\
 &  & +(\delta_{ij}-\frac{1}{3}R_{ikjl}x^{k}x^{l})dx^{i}dx^{j},\label{eq:metric}
\end{eqnarray}
where $t$ is the proper time measured by the observer centered at
the origin and $R_{\mu\nu\lambda\rho}\sim\frac{1}{\mathcal{R}^{2}}$
is the Riemann curvature tensor evaluated along the observer's world
line. Einstein summation convention is adopted.


As mentioned, for test masses in metric theories (including general relativity), to
embody WEP or the universality of free-falling motions, a separate
postulate of shortest paths (least action) is needed
\begin{equation}
S =\mu\int\sqrt{-g_{\mu\nu}\frac{dx^{\mu}}{dt}\frac{dx^{\nu}}{dt}}dt=\mu\int d\tau,\label{eq:action}
\end{equation}
where $\tau$ is the proper time experienced by the test mass $\mu$.
With the low energy conditions $v^{i}=\frac{dx^{i}}{dt}\ll 1$, $\frac{d\tau}{dt}\sim1$,
the metric component $g_{00}$ in the above action weights the most
in evaluating the length of the world line, and
therefore affects the motion mostly. Given the metric expansion in
Eq. (\ref{eq:metric}), the equation of free-falling or geodesic motions from the above
action can be cast into the simple and linearized form in local
Fermi normal frames \cite{Misner1973,Ni1978}
\begin{equation}
a^{i}=\frac{d^{2}x^{i}}{dt^{2}}=-R_{0j0i}x^{j}+\mathcal{O}(\frac{|x|v}{\mathcal{R}^{2}}),\label{eq:classical_geodesic}
\end{equation}
here $R_{0i0j}$ can be viewed as a symmetric matrix (the tidal matrix) with indices
$i,j=1,2,3$. As expected, this equation of motion is independent
of the internal structures and compositions of the classical test
particles or bodies, and therefore manifests the fulfillment of WEP
that viewed from the local frame of a free observer.

\section{Quantum wave packets in weakly curved spacetime}
Within the local Fermi normal frame as the common stage, the classical
model of test mass $\mu$ is replaced with the corresponding quantum wave packet
$\Psi(t,\vec{x})$ (usually a Gaussian wave packet), whose characteristic
scale $L$ is much smaller compared with the curvature radius $\mathcal{R}$
in the local frame. According to Eq. (\ref{eq:metric}), the metric
is then expanded in terms of the components of the curvature
tensor about a flat background. Without loss of generality, the wave
packet can be expanded in terms of Fourier modes
\begin{equation}
\Psi(t,\vec{x})=\frac{1}{(2\pi)^{\frac{3}{2}}}\int A(\vec{k})e^{-i\left(\omega(k) t-\vec{k}\cdot\vec{x}\right)}d^{3}k.\label{eq:Fourier}
\end{equation}
Normally, how does such quantum system respond to or interact with spacetime
curvatures could be answered by assuming a specific action
(or a Hamiltonian) that coupled to the gravitational field \cite{Birrell1984,Wald1994},
as for the case of classical test particles or bodies.
Generally, different choices may yield different dynamical equations.
For example, for the scalar wave $\Psi(t,\vec{x})$ considered
here, the minimal coupling may be a preferred choice since with this
choice the strong version of equivalence principle could be fulfilled.
Here we discard such priori given actions, and try to derive the preferred
mean motions of quantum test masses driven by the fundamental de Broglie
relations.

According to de Broglie relations, the total energy $\mathcal{E}$
and momentum $\vec{p}$ of the quantum system $\Psi(t,\vec{x})$
are associate to its frequency $\omega$ and wave vector $\vec{k}$.
Therefore, in the low energy limit 
that $p\ll\mu$, an
overall phase term can be factored out
\begin{equation}
\Psi(t,\vec{x})=e^{-2\pi i\mu t}\psi(t,\vec{x}),\label{eq:wave}
\end{equation}
where, according to Eq. (\ref{eq:Fourier}), one has
\[
\psi(t,\vec{x})=\frac{1}{(2\pi)^{\frac{3}{2}}}\int A(\vec{k})e^{-i(\frac{k^{2}}{4\pi\mu}t-\vec{k}\cdot\vec{x})}d^{3}k.
\]
This is because for each Fourier mode one has
$\omega(k)/2\pi=\mathcal{E}(k)=\mu+\frac{k^{2}}{(2\pi){}^{2}2\mu}+...$
with the characteristic or mean momentum $<p>=<k/2\pi>\ \ll\mu$.
This is a well-known result in non-relativistic quantum mechanics,
and usually the overall phase factor $e^{-2\pi i\mu t}$ is ignored
since it is expected to have no measurable effect in the low-energy
regime. But, concerning the universality of the free-falling motions
in curved spacetime, such phase factor from the rest mass $\mu$ will
play an important part in restoring the classical free-falling motions (given in Eq.
(\ref{eq:classical_geodesic})) or
the classical WEP that is viewed in local frames.
\begin{figure}
	\centering
\includegraphics[scale=0.45]{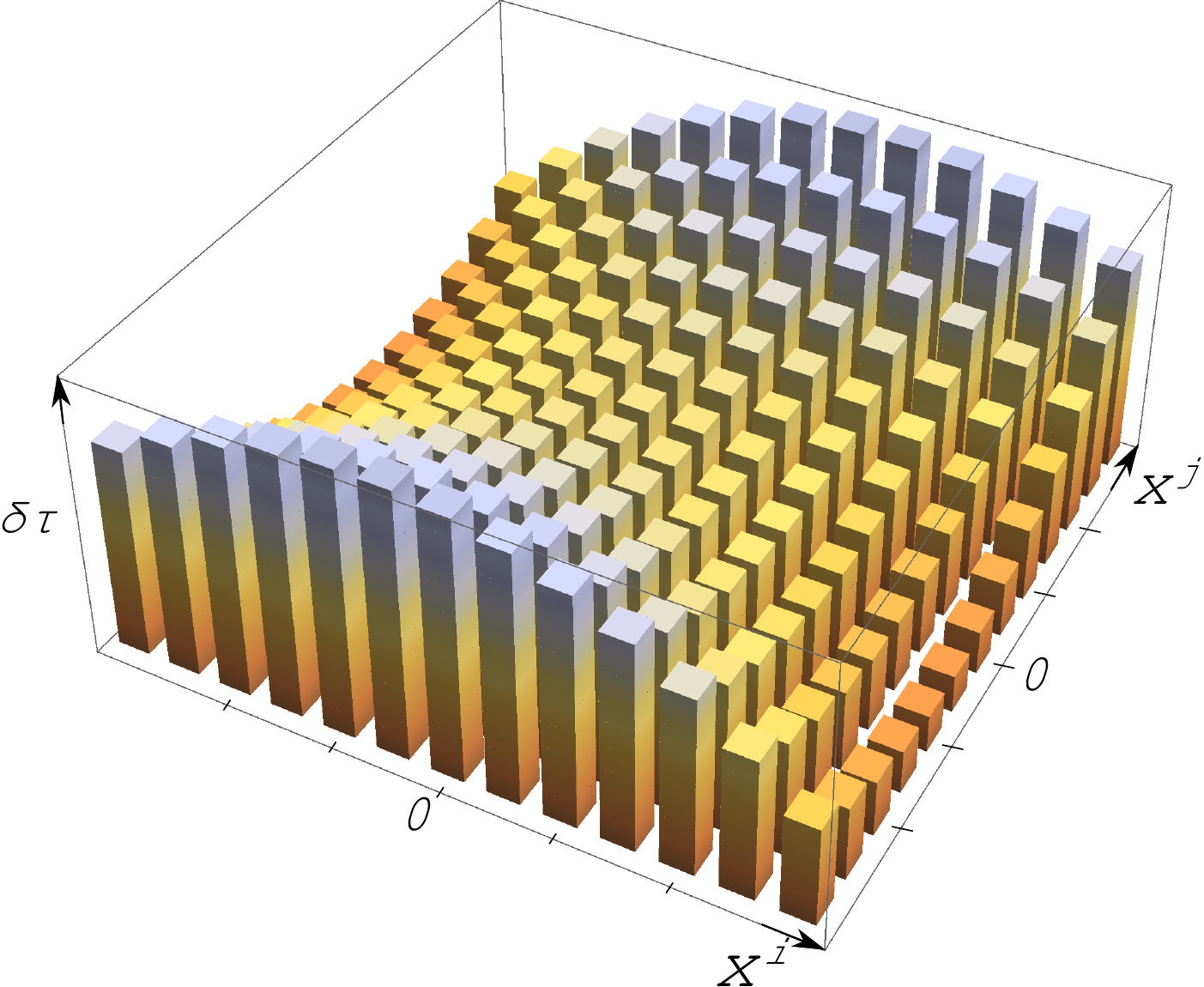}
\caption{In a general local Fermi normal frame, the histogram illustrates the
amount of proper time $\delta\tau=\sqrt{1+R_{0i0j}x^{i}x^{j}}\delta t$
counted at different points in a 2-dimensional space slice $x^{i}-x^{j}$
after the proper time lapse $\delta t$ recorded by the observer centered
at the origin. }
\label{fig:timeflow}
\end{figure}

Let us take the atomic clocks \cite{Cacciapuoti2009,Leveque2015} in gravitational
red shift experiments \cite{Will2014,Jetzer2017} and Bose-Einstein condensates in
free-falling experiments \cite{Zoest2010} as typical examples concerning
quantum systems in gravitational fields, compared to the curvature radius
the corresponding quantum systems or wave packets
are viewed as being sharply peaked at or
bounded to some small space regions.
It is assumed that such quantum systems are evolved according to the proper time
of the co-moving frame attached to the mass center of the experimental platform. While,
this is still of an averaged or approximated treatment. In
\cite{Zych2011,Zych2012,Pikovski2015,Pikovski2017},
such treatment is generalized for the case that the propagation of
a quantum system is in superpositions of different paths in gravitational
fields that are having different proper time dilations. In this work, with one
step forward, we suggest that with a natural
generalization more detailed behaviors of an extensively
distributed quantum wave could be studied. According to the classical free-falling motions in the
low-energy and weak field limit, we notice from Eq. (\ref{eq:metric})
- (\ref{eq:classical_geodesic}) that the test mass probes the variations
and non-uniformities of the metric component $g_{00}$ much stronger
than those of other components, since the test mass follows slowly (compared to the speed of light)
along a time-like worldline in the reference frame. The metric component $g_{00}$ can be
viewed as a measurement of the non-uniformity of the time flows or
clock rates at different points in the local Fermi normal frame, see Fig.
\ref{fig:timeflow} for an illustration. This naturally inspired us
that, for an extensive test body or system (quantum or classical)
moving slowly compared with the speed of light in gravitational field,
the time parameter of each part or at each point of the system should
be counted as the proper time $\tau$ measured by the clock located in that
part or at that point. In fact, for classical extensive bodies
such clock rate differences just give rise
to the tidal forces subjected to them from spacetime curvature, if the
priori given least action of matter fields coupled to gravity
is imposed. While, given
the de Broglie relations which indicate the associations between
4-momenta and spacetime periodic patterns, an important difference
between quantum and classical systems is that the phase at different
part of a quantum system is strongly correlated. Therefore,
without any priori given action, such differences
of the clock rates $d\tau/dt$ in the different parts of a
quantum wave packet $\Psi(\tau,\vec{x})$ in curved spacetime are expected
to generate additional phase variations and will drive the motions of the quantum
test system.

Without loss of generality, let us start with a space slice at $t=t_{0}$
in the local Fermi frame. After the coordinate time lapse $\delta t=t-t_{0}$,
which is the proper time lapse recorded by the observer centered at
the origin, the proper time lapse experienced at point $x^{i}$ will be
\begin{eqnarray}
\delta\tau & =&\sqrt{1+R_{0i0j}x^{i}x^{j}+\mathcal{O}(\frac{|x|^{2}v}{\mathcal{R}})}\delta t\nonumber \\
 & =&\left(1+\frac{1}{2}R_{0i0j}x^{i}x^{j}+\mathcal{O}(\frac{|x|^{2}v}{\mathcal{R}})\right)\delta t,\label{eq:tau}
\end{eqnarray}
see again Fig. \ref{fig:timeflow} for an illustration.
Therefore, the clock rate at point $x^{i}$ with respect to the observer's proper time reads
\begin{equation*}
\frac{\delta\tau}{\delta t}=1+\frac{1}{2}R_{0i0j}x^{i}x^{j}+\mathcal{O}(\frac{|x|^{2}v}{\mathcal{R}}).
\end{equation*}
For the wave
function $\Psi(\tau,\vec{x})$ distributed over the space slice, such
clock rate differences over the corresponding space region will generate
additional phase differences among the different parts of $\Psi(\tau,\vec{x})$
and then result in a variation in the corresponding wave vector.

\begin{figure}
	\centering
	\subfloat[]{\includegraphics[scale=0.3]{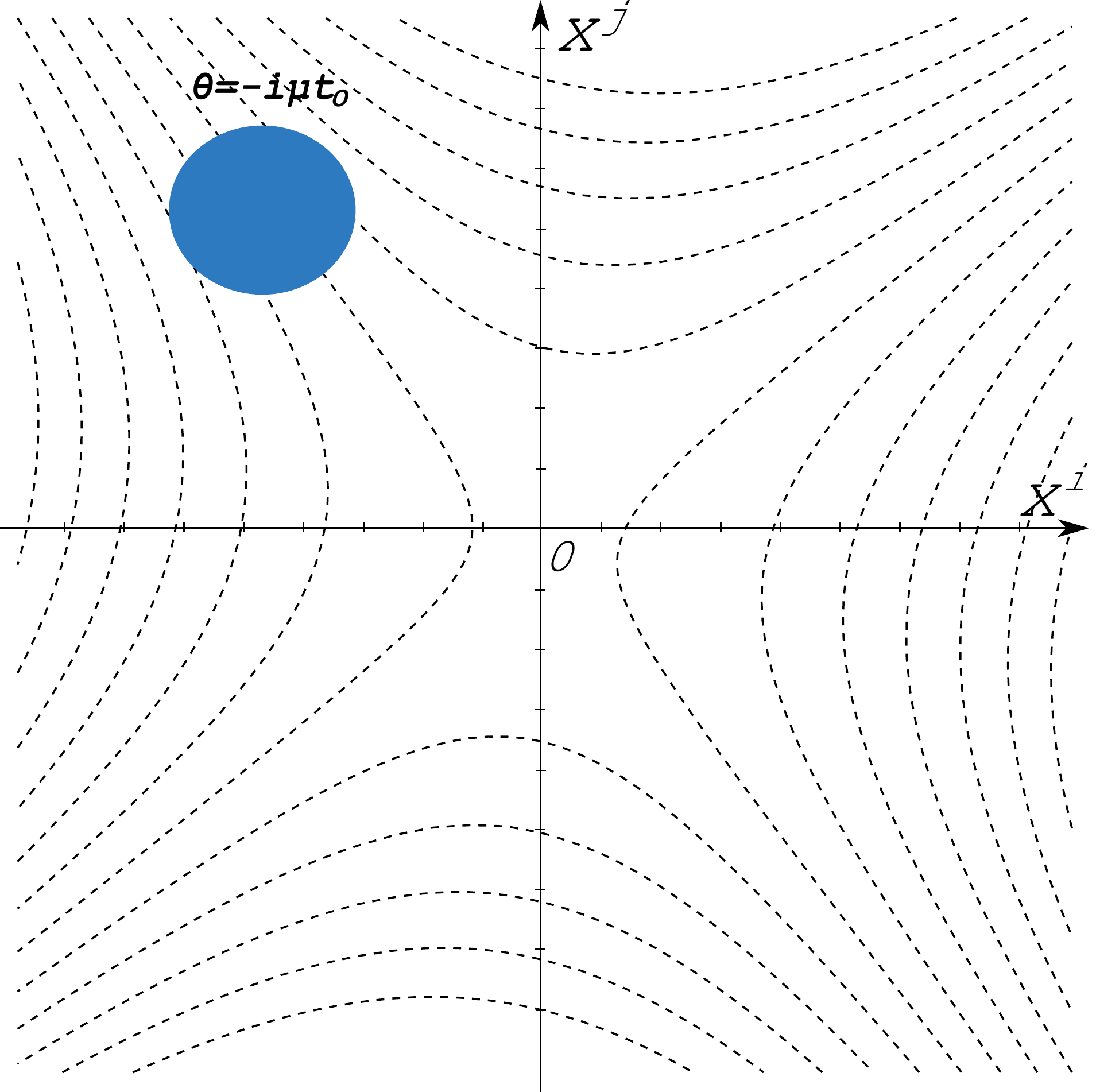}
		\label{subfig:phase_flat}}	
	\subfloat[]{\includegraphics[scale=0.3]{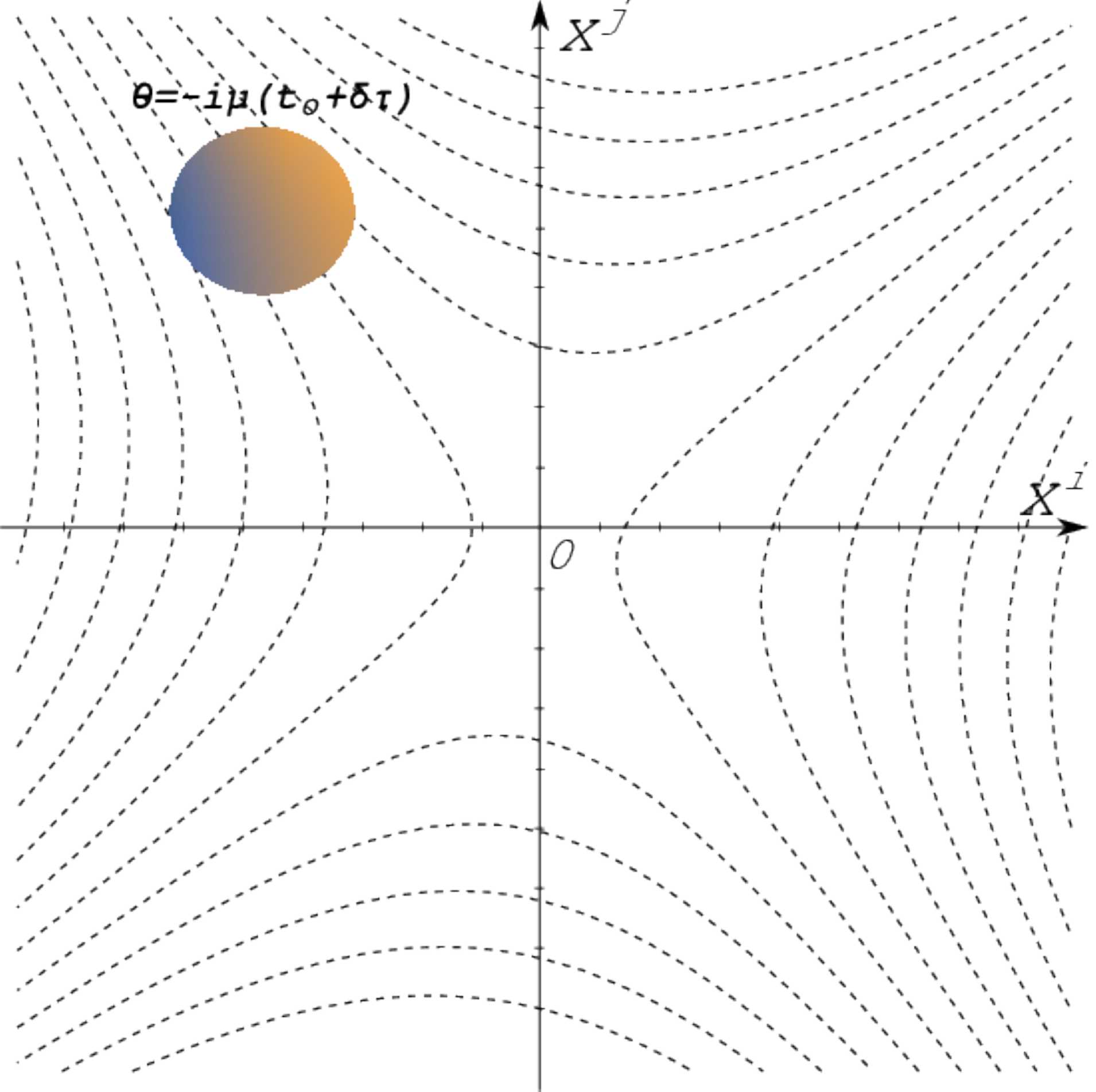}
		\label{subfig:phase_diff}}
	
	\subfloat[]{\includegraphics[scale=0.3]{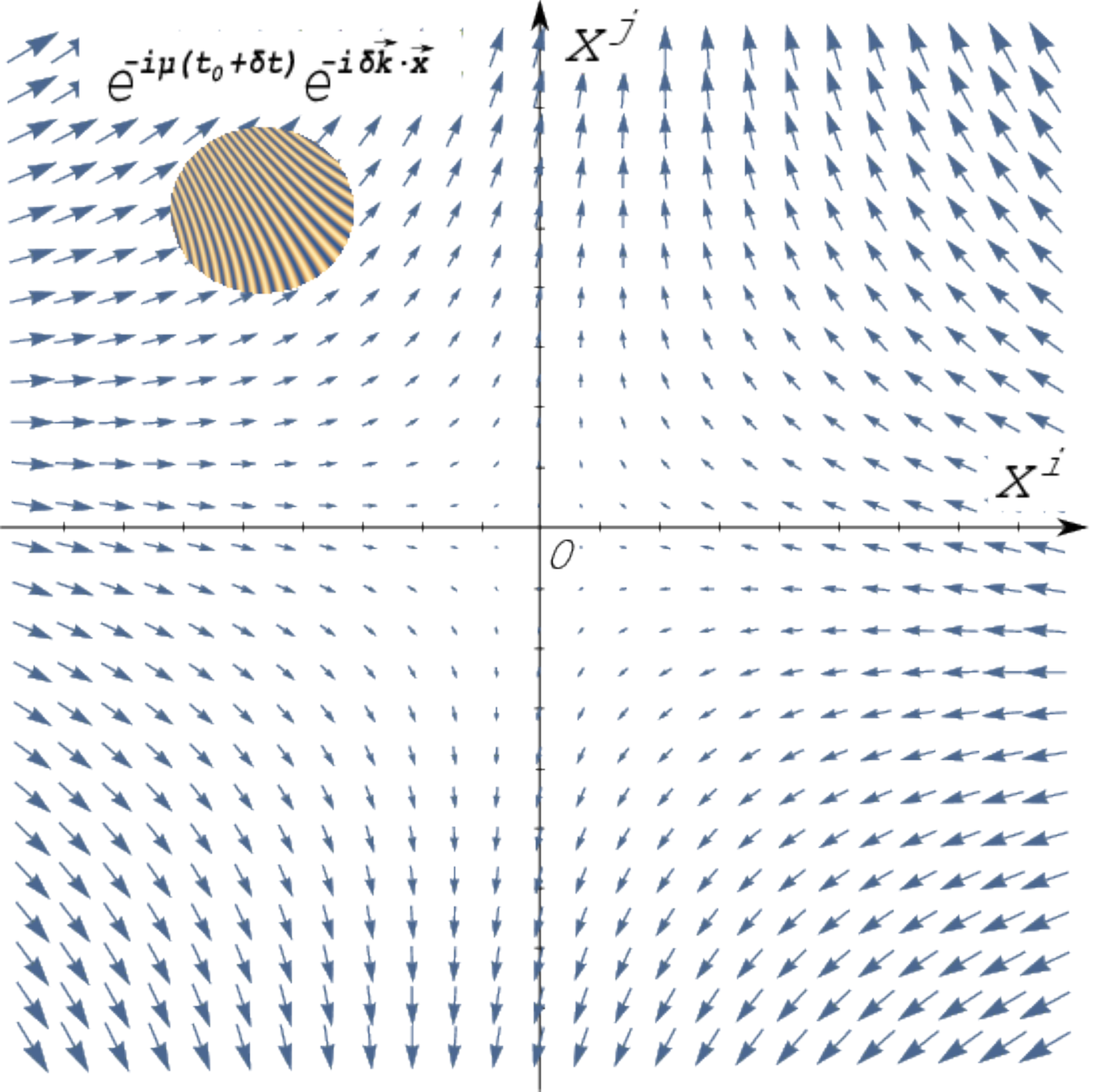}
		\label{subfig:phase_wave}}
	
	\caption{These illustrate the phase evolution of a Gaussian wave packet in
		a general Fermi normal frame in a weakly curved spacetime. (a): In
		a 2-dimensional space slice $x^{i}-x^{j}$ at $t=t_{0}$, the initial
		uniform phase in the overall factor $e^{-i\mu t_{0}}$ of the Gaussian
		wave packet is shown as a blue spot. The dashed contour lines mark
		the points with equal proper time lapse $\delta\tau$. (b): After
		the time lapse $\delta t$ counted by the observer that centered at the
		origin, the system is evolved into the subsequent space slice at $t=t_{0}+\delta t$.
		Due to the clock rate differences produced by spacetime curvature part $R_{0i0j}$,
		a phase difference is generated over the Gaussian wave packet. (c):
		Since phase variations in quantum waves are modulated to $2\pi$,
		the phase difference in Fig. (b) will be wrapped into a ripple over
		the wave packet. The mean increment in the wave vector $\delta\vec{k}$
		or momentum $\delta \vec{p}=\mu\delta \vec{v}=\delta \vec{k}/2\pi$ agrees with the
		tidal acceleration, as shown in form of
		a vector field, from the spacetime curvature part $R_{0i0j}$.}
	\label{fig:phase}
\end{figure}

In the space slice $t=t_{0}$, one denotes the wave function as $\Psi(t_{0},\vec{x})=e^{-2\pi i\mu t_{0}}\psi(t_{0},\vec{x})$
with a uniform initial phase $2\pi\mu t_{0}$ in the overall phase
factor from the rest mass $\mu$, and the possible initial non-uniformity
in the phase is left in the $\psi(t_{0},\vec{x})$ part,
see Fig. \ref{subfig:phase_flat} for illustration.
After the coordinate
time lapse $\delta t=t-t_{0}$, the wave function becomes $\Psi(t_{0},\vec{x})\rightarrow\Psi(t_{0}+\delta\tau,\vec{x})$
with $\delta\tau$ given in Eq. (\ref{eq:tau}), that
\begin{eqnarray}
&&  \Psi(t_{0}+\delta\tau,\vec{x})\nonumber \\
 & = & e^{-2\pi i\mu(t_{0}+\delta\tau)}\psi(t_{0}+\delta\tau,\vec{x})\nonumber \\
 & = & e^{-2\pi i\mu\left(\frac{1}{2}R_{0i0j}x^{i}x^{j}+\mathcal{O}(\frac{|x|^{2}v}{\mathcal{R}})\right)\delta t}e^{-i\mu(t_{0}+\delta t)}\psi(t_{0}+\delta\tau,\vec{x}),\nonumber \\
 & = & e^{-\pi i\mu R_{0i0j}x^{i}x^{j}\delta t}\Psi(t,\vec{x}).\label{eq:deltawave}
\end{eqnarray}
Therefore we have $\Psi(t_{0},\vec{x})\rightarrow e^{-\pi i\mu R_{0i0j}x^{i}x^{j}\delta t}\Psi(t,\vec{x})$,
where $\Psi(t,\vec{x})$ is obtained by time shifting the free wave
$\Psi(t_{0},\vec{x})$ with the common coordinate lapse $\delta t$
\[
\Psi(t,\vec{x})=e^{-2\pi i\mu t}\frac{1}{(2\pi)^{\frac{3}{2}}}\int A(\vec{k})e^{-i(\frac{k^{2}}{4\pi\mu}t-\vec{k}\cdot\vec{x})}d^{3}k.
\]
For the last equal in Eq. (\ref{eq:deltawave}), we make use of the low-energy
condition $\mu\gg\ <p>=<k>/2\pi$ and ignore the higher order terms
of $\mathcal{O}(\frac{<k>^{2}}{\mu^{2}})\mathcal{O}(\frac{|x|^{2}}{\mathcal{R}}\mu)$
in the phase. This could be made more clear when one considers the
Fourier expansions given in Eq. (\ref{eq:Fourier})
\begin{eqnarray*}
&&\Psi(t_{0}+\delta\tau,\vec{x})\nonumber\\
 & =&e^{-2\pi i\mu\left(\frac{1}{2}R_{0i0j}x^{i}x^{j}+\mathcal{O}(\frac{|x|^{2}v}{\mathcal{R}})\right)\delta t}e^{-2\pi i\mu(t_{0}+\delta t)}\frac{1}{(2\pi)^{\frac{3}{2}}}\int A(\vec{k})e^{-i\left[\frac{k^{2}}{4\pi\mu}\left(t_{0}+[1+\frac{1}{2}R_{0i0j}x^{i}x^{j}+\mathcal{O}(\frac{|x|^{2}v}{\mathcal{R}})]\delta t\right)-\vec{k}\cdot\vec{x}\right]}d^{3}k\\
 & =&e^{-2\pi i\mu\left(\frac{1}{2}R_{0i0j}x^{i}x^{j}+\mathcal{O}(\frac{|x|^{2}v}{\mathcal{R}})\right)\delta t}e^{-2\pi i\mu(t_{0}+\delta t)}\frac{1}{(2\pi)^{\frac{3}{2}}}\int A(\vec{k})e^{-i\left(\frac{k^{2}}{4\pi\mu}(t_{0}+\delta t)-\vec{k}\cdot\vec{x}\right)-i\mathcal{O}(\frac{k^{2}}{\mu^{2}})\mathcal{O}(\frac{|x|^{2}}{\mathcal{R}}\mu)\delta t}d^{3}k\\
 & =&e^{-\pi i\mu R_{0i0j}x^{i}x^{j}\delta t}\Psi(t,\vec{x}).
\end{eqnarray*}
Therefore, evolved to the subsequent space slice $t=t_{0}+\delta t$,
the clock rate differences in the weakly curved spacetime produce
a phase difference $\delta\theta=-\pi\mu R_{0i0j}x^{i}x^{j}\delta t$
in the wave function, see Fig. \ref{subfig:phase_diff} for an illustration.
Between two adjacent points $x^{i}$ and $x^{i}+\Delta x^{i}$ in
the quantum wave packet, such phase difference reads
\begin{equation}
\Delta(\delta\theta)=-2\pi\mu R_{0i0j}x^{i}\Delta x^{j}\delta t,\label{eq:dphase}
\end{equation}
Phase variations are modulated to $2\pi$, therefore the above phase
difference accumulated during $\delta t$ over the space interval
$\Delta\vec{x}$ will be wrapped into a ripple with the wave length
$2\pi\Delta x/\Delta(\delta\theta)$, see Fig. \ref{subfig:phase_wave}
for an illustration, and then results in a variation in the wave
vector $\delta\vec{k}$ after $\delta t$
\begin{equation}
\delta k_{i}=\frac{\Delta(\delta\theta)}{\Delta x^{i}}=-2\pi\mu R_{0j0i}x^{j}\delta t.\label{eq:dk}
\end{equation}
Now, according to the de Broglie relations $p_{i}=k_{i}/2\pi$, the
variations in wave vector imply the variations in the momenta $\delta p_{i}=\mu\delta v^{i}$,
therefore the mean value of the velocity variations or the acceleration
of the quantum test mass can be obtained as
\begin{eqnarray}
<\frac{\delta v^{i}}{\delta t}> & =&<\frac{\delta k_{i}}{2\pi\mu\delta t}>\nonumber \\
<a^{i}> & =&-R_{0j0i}<x^{j}>.\label{eq:quantum_geodesic}
\end{eqnarray}
This gives rise to the expected semi-classical version of the free-falling
or geodesic equation in gravitational fields viewed in local Fermi normal frames,
which is independent of the mass and composition of the quantum test system and will
restore Eq. (\ref{eq:classical_geodesic}) as one takes the classical limits. Therefore, within the
low energy regime, the classical WEP could be restored from such universality of
quantum mean motions in gravitational fields.

In the following, we give a more straightforward derivation of Eq. (\ref{eq:quantum_geodesic}).
Given the de Broglie relations, the velocity density takes the form
\begin{equation*}
v^{i}=\frac{p_{i}}{\mu}=-\frac{i}{4\pi\mu}(\Psi^{*}\nabla_{i}\Psi-\Psi\nabla_{i}\Psi^{*}).
\end{equation*}
According to the Fourier expansions of the quantum wave in Eq. (\ref{eq:Fourier})
and (\ref{eq:deltawave}), for the initial slice $t=t_{0}$ the mean velocity of the quantum test mass reads
\begin{eqnarray*}
&&<v^{i}(t_{0})>\\
&= & \int-\frac{i}{2\mu}\left(\Psi^{*}(t_{0},\vec{x})\nabla_{i}\Psi(t_{0},\vec{x})-\Psi(t_{0},\vec{x})\nabla_{i}\Psi^{*}(t_{0},\vec{x})\right)d^{3}x\\
&= & \frac{1}{2\pi\mu}\int|A(\vec{k})|^{2}k_{i}d^{3}k\\
&= & \frac{<k^{i}>}{2\pi\mu},
\end{eqnarray*}
and, as evolved to the subsequent slice $t=t_{0}+\delta t$, one has
\begin{eqnarray*}
  &&<v^{i}(t_{0}+\delta t)>\\
&= & \int-\frac{i}{2\mu}[e^{\pi i\mu R_{0i0j}x^{i}x^{j}\delta t}\Psi^{*}(t,\vec{x})\nabla_{i}\left(e^{-\pi i\mu R_{0i0j}x^{i}x^{j}\delta t}\Psi(t,\vec{x})\right)\\
&& -e^{-\pi i\mu R_{0i0j}x^{i}x^{j}\delta t}\Psi(t,\vec{x})\nabla_{i}\left(e^{\pi i\mu R_{0i0j}x^{i}x^{j}\delta t}\Psi^{*}(t,\vec{x})\right)]d^{3}x\\
&= & \frac{1}{2\pi\mu}\int|A(\vec{k})|^{2}k_{i}d^{3}k-R_{0k0i}\delta t\int\psi^{*}(t,\vec{x})x^{k}\psi(t,\vec{x})d^{3}x\\
&= & \frac{<k_{i}>}{2\pi\mu}-R_{0j0i}<x^{j}>\delta t \\
&& +\mathcal{O}(\frac{<x><v>}{\mathcal{R}}\mu)\delta t+\mathcal{O}(\frac{<k>^{2}}{\mu^{2}})\mathcal{O}(\frac{<x>}{\mathcal{R}}\mu)\delta t.
\end{eqnarray*}
With the increment in the mean velocity after $\delta t$
\begin{eqnarray}
<\delta v^{i}>&= & <v^{i}(t_{0}+\delta t)>-<v^{i}(t_{0})>\nonumber \\
&= & -R_{0j0i}<x^{j}>\delta t\nonumber\\
&&+\mathcal{O}(\frac{<x><v>}{\mathcal{R}}\mu)\delta t
+\mathcal{O}(\frac{<k>^{2}}{\mu^{2}})\mathcal{O}(\frac{<x>}{\mathcal{R}}\mu)\delta t,\label{eq:quantumdv}
\end{eqnarray}
we have
\begin{eqnarray}
<a^{i}>&=&\frac{<\delta v^i>}{\delta t}\nonumber\\
&= & -R_{0k0i}<x^{k}>\nonumber\\
&&+\mathcal{O}(\frac{<x><v>}{\mathcal{R}}\mu) +\mathcal{O}(\frac{<k>^{2}}{\mu^{2}})\mathcal{O}(\frac{<x>}{\mathcal{R}}\mu),\label{eq:quantuma}
\end{eqnarray}
which restores the expected semi-classical equation (\ref{eq:quantum_geodesic}).
One notices that this
derivation applies to general wave functions $\Psi(t, \vec{x})$,
and the resulting mean values (the first moments) $<\delta v^i>$
and $<a^i>$ do not depend on the structures or ``shapes'' of the quantum wave packets.

\section{Discussion}

Within the low-energy realm, if one takes into account
the clock rate differences or gravitational time dilations in the different parts of a test quantum
wave packet that is distributed over a weakly curved spacetime, the de
Broglie relations for the correlated phase in the quantum wave will
naturally drive the mean motions of the quantum test system to follow the classical free-falling trajectories.
The mean velocity variation given in Eq. (\ref{eq:quantumdv}) or
the mean acceleration in Eq. (\ref{eq:quantuma}) in weak gravitational fields does not
depend on the masses or compositions of the quantum test systems or
the shapes of the corresponding quantum waves.
Such mean motions of free quantum test systems in gravitational fields
agree exactly with the classical free-falling or geodesic
motions given in Eq. (\ref{eq:classical_geodesic}), which are determined by the background geometries.
Therefore, as one takes the classical limit, such universality of quantum mean motions implies the
WEP in the macroscopic world. Remember again that the classical free-falling or
geodesic equation in geometric (metric) theories of gravitation is
obtained by an additional postulate of shortest paths for test masses
or a universal coupling of matter fields to gravity,
while here the quantum equation does not rely on any such priori given
action or Hamiltonian but can be deduced naturally with the aids of the de Broglie relations.
This reveals, in a new angle, the connections between the fundamental
quantum laws and the general notion of
curved spacetime, and also a possible link between the de Broglie relations and WEP.
More interestingly, this also suggests a novel perspective that WEP,
as the foundation of gravitational theories, could be deeply rooted in quantum physics
and be emerged from the quantum world.

Only mean motions of test quantum wave packets are considered here.
Quantum fluctuations beyond the mean motions
that are affected by weakly curved spacetime can be studied
in terms of variances, covariances and higher moments within this formalism.
For wave packets that deviate largely from a sharply
peaked Gaussian packet (classical limits) or are Gaussian packets with highly asymmetric shapes,
whether or not quantum fluctuations affected or caused by spacetime curvatures
could produce measurable effects is of interest and worthy of further investigations.
At last, we have made use of Fermi normal frames for the low-energy
weak-field regime and the linearized classical geodesic equation to bring together the quantum and gravitational effects
to interweave with each other. To further reinforce the central idea, extensions to
more general cases, especially with internal degrees of freedom such as spins and
couplings to strong gravitational fields, could be addressed.

\textbf{Acknowledgments}.---
The authors thank Yao Cheng from Tsianghua University and Yun Kau
Lau from Chinese Academy of Sciences for the valuable suggestions
and discussions. The NSFC Grands No. 11571342, National Key R\&D Program of China
No.2017YFC0602202 and Natural Science Basic Research Plan in Shaanxi Province of China No.
2017JQ1028 are acknowledged. The authors are also supported by the State Key
Laboratory of Scientific and Engineering Computing, Chinese Academy of Sciences.

\end{document}